\input amstex

%%% ====================================================================
%%%  This is a documentstyle for use with AMSTeX  2.x,
%%%  in production of the English translation of
%%%  the journal ``Mathematical Notes.''
%%%  It makes use of the documentstyle amsppt and adds
%%%  some additional features.
%%% ====================================================================

\catcode`\@=11

\input amsppt.sty \relax

\pagewidth{29pc}

\outer\def\ahead#1\endahead%
{\removelastskip\vskip18pt plus2pt minus2pt\csname head\endcsname\uppercase{#1}\endhead\nobreak}
\outer\def\bhead#1\endbhead%
{\removelastskip\vskip15pt plus1pt minus1pt\csname head\endcsname{\it #1}\endhead\nobreak}
\outer\def\chead#1\endchead%
{\medskip\csname subhead\endcsname#1\endsubhead\nobreak}
\outer\def\dhead#1\enddhead%
{\medskip\csname subsubhead\endcsname#1\endsubsubhead\nobreak}

\def\demo{\csname definition\endcsname}
\def\enddemo{\enddefinition}

\def\remark{\csname definition\endcsname}
\def\endremark{\enddefinition}

\def\example{\csname definition\endcsname}
\def\endexample{\enddefinition}

\def\translator{%
  \let\savedef@\translator
  \def\translator##1\endtranslator{\let\translator\savedef@
    \edef\thetranslator@{\noexpand\nobreak\noexpand\medskip
      \noexpand\line{\noexpand\ninepoint\hfil
      \frills@{Translated by }{##1}}%
       \noexpand\nobreak}}%
  \nofrillscheck\translator}

%%%%\Monograph
\monograph@false
\let\headmark\eat@

\begingroup
  
\let\ahead\relax \let\bhead\relax \let\chead\relax

\gdef\widestnumber{\begingroup
  \let\ahead\relax \let\bhead\relax \let\chead\relax
  \expandafter\endgroup\setwidest@}
\gdef\setwidest@#1#2{%
   \ifx#1\ahead\setbox\tocheadbox@\hbox{#2.\enspace}%
   \else\ifx#1\bhead\setbox\tocsubheadbox@\hbox{#2.\enspace}%
   \else\ifx#1\chead\setbox\tocsubheadbox@\hbox{#2.\enspace}%
   \else\ifx#1\key\refstyle A%
       \setboxz@h{\refsfont@\keyformat{#2}}%
       \refindentwd\wd\z@
   \else\ifx#1\no\refstyle C%
       \setboxz@h{\refsfont@\keyformat{#2}}%
       \refindentwd\wd\z@
   \else\ifx#1\page\setbox\z@\hbox{\quad\bf#2}%
       \pagenumwd\wd\z@
   \else\ifx#1\item
       \edef\next@{\the\revert@\rosteritemwd\the\rosteritemwd\relax
              \revert@{\the\revert@}}%
       \revert@\expandafter{\next@}%
       \setboxz@h{(#2)}\rosteritemwd\wdz@
   \else\message{\string\widestnumber\space not defined for this
      option (\string#1)}%
\fi\fi\fi\fi\fi\fi\fi}

\endgroup

\mathsurround=1.2pt

\def\punctcheck@{\futurelet\nextchar@\removemathsurround@}

\let\everymath@\everymath
\let\everydisplay@\everydisplay

\newtoks\everymath
\everymath@{\the\everymath}

\newtoks\everydisplay
\everydisplay@{\the\everydisplay}

\everymath{\aftergroup\punctcheck@}

\def\removemathsurround@{%
  {\edef\next{\noexpand\nextchar@}%
  \ifcat.\noexpand\nextchar@
    \expandafter\ifx\next,\else\expandafter\ifx\next.\else
      \expandafter\ifx\next;\else\kern-\mathsurround\fi\fi\fi
  \fi
  \let\nextchar@\relax
  }%
}

\let\@@eqno=\eqno

\let\@@leqno=\leqno

\def\eqno{\everymath{}\@@eqno}
\def\leqno{\everymath{}\@@leqno}

\def\mathcomma@{\mathpunct{\mkern1.2mu\mathchar"013B\mkern1.2mu}}%
\def\mathsemicolon@{\mathpunct{\mkern3mu\mathchar"003B\mkern3mu}}%
\def\mathcolon@{\mathrel{\mkern\thickmuskip\nonscript\mkern-\thickmuskip
  \mathchar"003A\mkern\thickmuskip\nonscript\mkern-\thickmuskip}}

\mathcode`\,="8000 \mathcode`\;="8000 \mathcode`\:="8000

\begingroup
\catcode`\,=\active \catcode`\;=\active \catcode`\:=\active
\global\everymath@{%
  \let,\mathcomma@ \let;\mathsemicolon@ \let:\mathcolon@
  \the\everymath}
\global\everydisplay@{%
  \let,\mathcomma@ \let;\mathsemicolon@ \let:\mathcolon@
  \def\.{\thinspace.}%
  \the\everydisplay}
\endgroup

\def\cprime{\/\begingroup\everymath{}\m@th$'$\snug\hskip-0.05em\endgroup}
 % ``myagkii znak'' synonym for \cprime

\font@\ninerm=cmr9
\font@\ninei=cmmi9    \skewchar\ninei='177
\font@\ninesy=cmsy9   \skewchar\ninesy='60
\font@\nineex=cmex9
\font@\ninebf=cmbx9
\font@\nineit=cmti9
\font@\ninesmc=cmcsc9
\font@\ninesl=cmsl9

\font@\ninemsa=msam9
\font@\ninemsb=msbm9
\font@\nineeufm=eufm9

\newtoks\ninepoint@
\def\ninepoint{\normalbaselineskip11\p@
 %\abovedisplayskip10\p@ plus2.4\p@ minus7.2\p@
 \abovedisplayskip10\p@ plus2.4\p@ minus4.2\p@
 \belowdisplayskip\abovedisplayskip
 \abovedisplayshortskip\z@ plus2.4\p@
 \belowdisplayshortskip5.6\p@ plus2.4\p@ minus3.2\p@
 \textonlyfont@\rm\ninerm \textonlyfont@\it\nineit
 \textonlyfont@\bf\ninebf \textonlyfont@\smc\ninesmc
 \textonlyfont@\sl\ninesl
 \ifsyntax@\def\big##1{{\hbox{$\left##1\right.$}}}%
  \let\Big\big \let\bigg\big \let\Bigg\big
 \else
  \textfont\z@\ninerm \scriptfont\z@\sevenrm
       \scriptscriptfont\z@\fiverm
  \textfont\@ne\ninei \scriptfont\@ne\seveni
       \scriptscriptfont\@ne\fivei
  \textfont\tw@\ninesy \scriptfont\tw@\sevensy
       \scriptscriptfont\tw@\fivesy
  \textfont\thr@@\nineex \scriptfont\thr@@\sevenex
   \scriptscriptfont\thr@@\sevenex
  \textfont\itfam\nineit \scriptfont\itfam\sevenit
   \scriptscriptfont\itfam\sevenit
  \textfont\bffam\ninebf \scriptfont\bffam\sevenbf
   \scriptscriptfont\bffam\fivebf
  \textfont\msafam\ninemsa \scriptfont\msafam\sevenmsa
   \scriptscriptfont\msafam\fivemsa
  \textfont\msbfam\ninemsb \scriptfont\msbfam\sevenmsb
   \scriptscriptfont\msbfam\fivemsb
  \textfont\eufmfam\nineeufm \scriptfont\eufmfam\seveneufm
   \scriptscriptfont\eufmfam\fiveeufm
 \setbox\strutbox\hbox{\vrule height7.5\p@ depth3.5\p@ width\z@}%
 \setbox\strutbox@\hbox{\raise.5\normallineskiplimit\vbox{%
   \kern-\normallineskiplimit\copy\strutbox}}%
 \setbox\z@\vbox{\hbox{$($}\kern\z@}\bigsize@1.2\ht\z@
 \fi
 \normalbaselines\ninerm\dotsspace@1.5mu\ex@.2326ex\jot3\ex@
 \the\ninepoint@}

\font@\twelverm=cmr10 scaled\magstep1
\font@\twelvei=cmmi10 scaled\magstep1    \skewchar\twelvei='177
\font@\twelvesy=cmsy10 scaled\magstep1   \skewchar\twelvesy='60
\font@\twelveex=cmex10 scaled\magstep1
\font@\twelvebf=cmbx10 scaled\magstep1
\font@\twelveit=cmti10 scaled\magstep1

\font@\twelvemsa=msam10 scaled\magstep1
\font@\twelvemsb=msbm10 scaled\magstep1
\font@\twelveeufm=eufm10 scaled\magstep1

\newtoks\twelvepoint@
\def\twelvepoint{\normalbaselineskip15\p@
 \abovedisplayskip10\p@ plus2.4\p@ minus7.2\p@
 \belowdisplayskip\abovedisplayskip
 \abovedisplayshortskip\z@ plus2.4\p@
 \belowdisplayshortskip5.6\p@ plus2.4\p@ minus3.2\p@
 \textonlyfont@\rm\twelverm \textonlyfont@\bf\twelvebf
 \textonlyfont@\it\twelveit
 \ifsyntax@\def\big##1{{\hbox{$\left##1\right.$}}}%
  \let\Big\big \let\bigg\big \let\Bigg\big
 \else
  \textfont\z@\twelverm \scriptfont\z@\tenrm
       \scriptscriptfont\z@\sevenrm
  \textfont\@ne\twelvei \scriptfont\@ne\teni
       \scriptscriptfont\@ne\seveni
  \textfont\tw@\twelvesy \scriptfont\tw@\tensy
       \scriptscriptfont\tw@\sevensy
  \textfont\thr@@\twelveex \scriptfont\thr@@\tenex
   \scriptscriptfont\thr@@\sevenex
  \textfont\itfam\twelveit \scriptfont\itfam\tenit
   \scriptscriptfont\itfam\sevenit
  \textfont\bffam\twelvebf \scriptfont\bffam\tenbf
   \scriptscriptfont\bffam\sevenbf
  \textfont\msafam\twelvemsa \scriptfont\msafam\tenmsa
   \scriptscriptfont\msafam\sevenmsa
  \textfont\msbfam\twelvemsb \scriptfont\msbfam\tenmsb
   \scriptscriptfont\msbfam\sevenmsb
  \textfont\eufmfam\twelveeufm \scriptfont\eufmfam\teneufm
   \scriptscriptfont\eufmfam\seveneufm
 \setbox\strutbox\hbox{\vrule height12\p@ depth5\p@ width\z@}%
 \setbox\strutbox@\hbox{\raise.5\normallineskiplimit\vbox{%
   \kern-\normallineskiplimit\copy\strutbox}}%
 \setbox\z@\vbox{\hbox{$($}\kern\z@}\bigsize@1.2\ht\z@
 \fi
 \normalbaselines\twelverm\dotsspace@1.5mu\ex@.2326ex\jot3\ex@
 \the\twelvepoint@}

\font@\svnteenrm=cmr10 scaled\magstep3
\font@\svnteeni=cmmi10 scaled\magstep3    \skewchar\svnteeni='177
\font@\svnteensy=cmsy10 scaled\magstep3   \skewchar\svnteensy='60
\font@\svnteenex=cmex10 scaled\magstep3
\font@\svnteenbf=cmbx10 scaled\magstep3
\font@\svnteenit=cmti10 scaled\magstep3

\font@\svnteenmsa=msam10 scaled\magstep3
\font@\svnteenmsb=msbm10 scaled\magstep3
\font@\svnteeneufm=eufm10 scaled\magstep3

\newtoks\svnteenpoint@
\def\svnteenpoint{\normalbaselineskip15\p@
 \abovedisplayskip10\p@ plus2.4\p@ minus7.2\p@
 \belowdisplayskip\abovedisplayskip
 \abovedisplayshortskip\z@ plus2.4\p@
 \belowdisplayshortskip5.6\p@ plus2.4\p@ minus3.2\p@
 \textonlyfont@\rm\svnteenrm \textonlyfont@\bf\svnteenbf
 \textonlyfont@\it\svnteenit
 \ifsyntax@\def\big##1{{\hbox{$\left##1\right.$}}}%
  \let\Big\big \let\bigg\big \let\Bigg\big
 \else
  \textfont\z@\svnteenrm \scriptfont\z@\tenrm
       \scriptscriptfont\z@\sevenrm
  \textfont\@ne\svnteeni \scriptfont\@ne\teni
       \scriptscriptfont\@ne\seveni
  \textfont\tw@\svnteensy \scriptfont\tw@\tensy
       \scriptscriptfont\tw@\sevensy
  \textfont\thr@@\svnteenex \scriptfont\thr@@\tenex
   \scriptscriptfont\thr@@\sevenex
  \textfont\itfam\svnteenit \scriptfont\itfam\tenit
   \scriptscriptfont\itfam\sevenit
  \textfont\bffam\svnteenbf \scriptfont\bffam\tenbf
   \scriptscriptfont\bffam\sevenbf
  \textfont\msafam\svnteenmsa \scriptfont\msafam\tenmsa
   \scriptscriptfont\msafam\sevenmsa
  \textfont\msbfam\svnteenmsb \scriptfont\msbfam\tenmsb
   \scriptscriptfont\msbfam\sevenmsb
  \textfont\eufmfam\svnteeneufm \scriptfont\eufmfam\teneufm
   \scriptscriptfont\eufmfam\seveneufm
 \setbox\strutbox\hbox{\vrule height12\p@ depth5\p@ width\z@}%
 \setbox\strutbox@\hbox{\raise.5\normallineskiplimit\vbox{%
   \kern-\normallineskiplimit\copy\strutbox}}%
 \setbox\z@\vbox{\hbox{$($}\kern\z@}\bigsize@1.2\ht\z@
 \fi
 \normalbaselines\svnteenrm\dotsspace@1.5mu\ex@.2326ex\jot3\ex@
 \the\svnteenpoint@}

\newbox\firstheadline

\output={\faaoutput}
\def\faaoutput{\shipout\vbox{\makeheadline\pagebody\makefootline}%
  \advancepageno
  \ifnum\outputpenalty>-\@MM \else\dosupereject\fi}

\def\pagebody{\advance\vsize-0.6\ht\firstheadline
\vbox to\vsize{\boxmaxdepth\maxdepth \pagecontents}}

\def\makeheadline{\vbox to\z@{\vskip-22.5\p@
  \line{\vbox to8.5\p@{}\the\headline}\vss}\nointerlineskip}

\def\makefootline{\baselineskip24\p@\line{\the\footline}}

\def\dosupereject{\ifnum\insertpenalties>\z@ % something is being held over
  \line{}\kern-\topskip\nobreak\vfill\supereject\fi}

\def\pagecontents{\ifvoid\topins\else\unvbox\topins\fi
  \dimen@=\dp\@cclv \unvbox\@cclv % open up \box255
  \ifvoid\footins\else % footnote info is present
    \vskip\skip\footins
    \footnoterule
    \unvbox\footins\fi
      \ifr@ggedbottom \kern-\dimen@ \vfil \fi}

\def\footnoterule{\vskip-2pt plus 1pt
  \hrule width 1.07in \kern 2.6\p@} % the \hrule is .4pt high

%==========================================
\def\leftheadline{{\ninerm\the\pageno} \hfill\ninepoint\lefttext
%%\ \nineit\ et al.
\hfill}
\def\rightheadline{\hfill\ninepoint\righttext
\hfill {\ninerm\the\pageno}}
%============================================
\headline{\ifvoid\firstheadline\ifodd\pageno
 \rightheadline\else\leftheadline\fi\else\box\firstheadline\fi}
%%%%%%%%%%%%%%%%%%%%%%%%%%%%%%%%%%%%%%%%%%%%%%%%%%%%%%%%%%%%%%%%%%%%%%
\outer\def\enddocument{\par% \par will do a runaway check for \endref
  \add@missing\endRefs
  \add@missing\endroster \add@missing\endproclaim
  \add@missing\enddefinition
  \add@missing\enddemo \add@missing\endremark \add@missing\endexample
 \ifmonograph@ % do nothing
 \else
 \nobreak
 \thetranslator@
 \count@\z@ \loop\ifnum\count@<\addresscount@\advance\count@\@ne
 \csname address\number\count@\endcsname
 \csname email\number\count@\endcsname
 \repeat
\fi
 \vfill\supereject\end}

\csname man.sty\endcsname
\pagewidth{35.59pc} \pageheight{50.98pc}
\magnification=1095
\hoffset=-0.17mm
\voffset=-3.7mm
\baselineskip=12pt plus 0.2pt minus 0.2pt
\abovedisplayskip=7pt plus 2pt minus 2pt
\belowdisplayskip=7pt plus 2pt minus 2pt
\abovedisplayshortskip=5pt plus 1pt minus 3pt
\belowdisplayshortskip=5pt plus 1pt minus 3pt

\def\titleinfo#1{\flushpar{\eightit }
   \vglue 3.89pc
\centerline{\svnteenpoint \bf #1}}

\def\titleinfobc#1{%\flushpar{\eightit }%
\vglue 1.11pc%
\centerline{\raise2pt\hbox to 44mm{%
\vbox{\hrule width 44mm\vskip2pt\hrule width 44mm}}%
\hfill\elvbf{BRIEF COMMUNICATIONS}\hfill%
\raise2pt\hbox to 44mm{%
\vbox{\hrule width 44mm\vskip2pt\hrule width 44mm}}%
}%
\vglue 2.14pc%
%\flushpar{\eightit }%additional line!!!
\centerline{\svnteenpoint \bf #1}}%

\def\extratitleline#1{\vskip6pt% minus1.5pt%
\centerline{\svnteenpoint\bf #1}}

\def\authorinfo#1#2{\bigskip\vskip2pt
\centerline{\elvbf{#1}}
\medskip
\centerline{\eightrm #2}}

%==================================================================
\newif\iffirstpage     \firstpagetrue
\output={\plainoutput}
                                                       %%put \pageno !!
\footline={\iffirstpage \global\firstpagefalse      %%\ifnum\pageno=281
\firstfootline
\else{\ifodd\pageno\rightfootline
\else\leftfootline\fi}\fi}

\def\firstfootline{\ifodd\pageno\rightfirstfootline
\else\leftfirstfootline\fi}

\def\rightfirstfootline{}
\def\leftfirstfootline{}
\def\rightfootline{}
\def\leftfootline{}

%===================================================================

\redefine\Refs{\bgroup\vskip10pt%
\centerline{REFERENCES}\nobreak\vskip7.17pt\leftskip5pt\frenchspacing
\ninepoint\nobreak}    %% 7.17pt instead of 10pt (counting in \medbreak)
\redefine\endRefs{\egroup\vskip12pt minus 1pt}
\def\authoraddress#1{\vskip-2pt%
\leftline{\ninesmc\ \ \ \ #1}}

\def\authoremail#1{\vskip-2pt%
\leftline{\ninepoint{\it\ \ \ \ E-mail\/}: #1}}

\def\ed--transl#1{\vskip10pt minus 1pt
\eightbf EDITOR: \
\underbar{$\hphantom{\text{Alex-Alexandrovskii}}$}\hfill
\eightrm Translated by #1}
\def\sted--transl#1{\vskip10pt minus 1pt
\eightbf STYLE EDITOR: \
\underbar{$\hphantom{\text{Alex-Alexandrovskii}}$}\hfill
\eightrm Translated by #1}
\def\bcskip{\vglue30pt\hrule width \hsize
\kern3pt\vglue30pt}
\TagsOnRight
\NoRunningHeads

\define\q{\quad}
\define\qq{\qquad}

\define\hph#1{\hphantom{#1}}

            %textmode

%%%\def\sserif{\bold}
.pk scaled 1000
.pk scaled 1000
.pk scaled 1000
%\font\sevenssf=cmss7.pk scaled 1000       NOT LOADABLE!
%\font\fivessf=cmss5.pk scaled 1000        NOT LOADABLE!
.pk scaled \magstep2
%\define\sserif#1{\mathchoice{\hbox{\tenssf #1}}{\hbox{\tenssf #1}}%
%{\hbox{\sevenssf #1}}{\hbox{\fivessf #1}}}

%%\redefine\ge{\geqslant}
%%\redefine\geq{\geqslant}
%%\redefine\le{\leqslant}
%%\redefine\leq{\leqslant}
\font\elevenbf=cmbx10 scaled 1095
\def\elvbf#1{\hbox{\elevenbf #1}}
\hyphenation{ap-prox-i-ma-tion ap-prox-i-ma-tions}

\pageno=1

%%\baselineskip24pt minus1.5pt
%-----------------------------------------------
\loadbold %% (not required if \boldkey or
%% \boldsymbol are not used!)
%-----------------------------------------------
\NoBlackBoxes %% do not uncomment!
%%%%%%%%%

% % % % % % % % % THE MACROS GROUP % % % % % % %
%% N.B. Please, check that the argument(s) %%
%% of "\operatorname(s)" (inside braces) %%
%% in the definitions below are correct! %%
%% (Note that all the macros must constitute
%% a solid group without empty lines.)

\def\0{\text{\bf 0}}

\def\1{\text{\bf 1}}

\def\supp{\text{\rm supp}}
\def\Tr{\text{\rm Tr}}

\def\otm{\operatornamewithlimits{\otimes}}
\def\cuplim{\operatornamewithlimits{\bigcup}}
%%%%%%%%%

%Journalinfo
% [PAGES]
%\def\Russpage{??}% [FOR A SINGLE-PAGE ARTICLE]

%%%%%%%%%

% % % % % % % % THE TITLE GROUP % % % % % % % %
\comment
\endcomment
\titleinfo%bc%
{Nondigital Implementation}
\extratitleline%
{of the Arithmetic of Real Numbers}
\vskip3pt %(editor's correction when needed)
\extratitleline%
{by Means of Quantum Computer Media}
\def\righttext{\uppercase{
Quantum Computer Media
}}
%%%%%%%%%

\def\lefttext{\uppercase{G.~L.~Litvinov, \ V.~P.~Maslov, \ G.~B.~Shpiz}}
\authorinfo{G.~L.~Litvinov, V.~P.~Maslov, and G.~B.~Shpiz}{}

\topmatter
\abstract\nofrills
\ninepoint
{\bf Abstract}---%[ABSTRACT]
 In the framework of a model for quantum computer media, a
nondigital implementation of the arithmetic of the real numbers
is described.
 For this model, an elementary storage ``cell'' is
an ensemble of qubits (quantum bits).
 It is found that to store
an arbitrary real number it is sufficient to use four of these
ensembles and the arithmetical operations can be implemented by
fixed quantum circuits.
\par\removelastskip\medskip\vskip1ex\noindent
{\eightsmc Key words}: \it %[ON A NEW LINE]
%[4--6 TERMS IN DESCENDING ORDER OF IMPORTANCE]
quantum media, quantum computer, arithmetic, qubit,
$q$-ensemble.
%----------------------------------------------
\endabstract
\endtopmatter
%%%%%%%%%%%%%%%%%%%%%%%%%%%%%%%%%%%%%%%%%%%%%%%

\document
\par\removelastskip\vskip18pt plus3pt minus3pt

\ahead{Introduction}\endahead

 In this note an implementation
of the arithmetic of real numbers is
described in the framework of a model for quantum computer
media (QCM).
 This model is an extension of the well-known
standard model for quantum computers.
 For such quantum computer
media an elementary storage ``cell'' is an ensemble of qubits (i.e.,
quantum bits).
 It is found that to store an arbitrary real number it is
sufficient to use four of these ensembles and the arithmetical
operations
can be carried out with a fixed number of elementary steps.
 Here
any number is represented in nondigital form.
 The representation of such a number in digital form
(e.g., in the form of a binary or decimal fraction)
is a separate problem of statistical estimation.
 Another approach to quantum computations
over continuous variable is presented, e.g., in \cite{1, 2}.

\ahead{1.
 Standard model of quantum computations}\endahead

 The idea of quantum computing was first put forward by P.~ Benioff
\cite{3, 4}, Yu.~I.~ Manin \cite{5}, R.~ P.~ Feynman \cite{6, 7}, and
A.~Peres \cite {8}.
 In Feynman's paper \cite{7} this idea was discussed
in detail.
 D.
 Deutsch \cite{9} stated a general formal definition of the
so-called quantum Turing machine.
 In \cite{10} he presented another
(equivalent but more convenient) model, which is now
regarded as standard.

 We shall recall some basic concepts and the corresponding
notation for a version of the standard model presented in
\cite{11--13} (see these papers for details, as well as, e.g.,
\cite{14}).

 Let
$X$
be a finite set.
 Denote by
$\Cal B(X)$
the set of all Boolean functions
defined on
$X$
and taking values 0 or 1.
 Let
$\Cal H(X)$
be the
complex Hilbert space with
$\Cal B(X)$
as an orthonormal basis;
so if
$X$
contains
$n$
elements, then the dimension of the space
$\Cal H(X)$
is
$2^n$.
 Let
${\bold L}(X)$
be the algebra of all linear operators
in
$\Cal H(X)$,
${\bold U}(X)$
the group of all unitary operators in
$\Cal H(X)$,
and
${\bold D}(X)$
the set of all density
operators, i.e., positive selfadjoint operators in
$\Cal H(X)$
whose trace is equal to 1.

 If
$\Delta$
is a set of bits forming the storage of a classical
computer, then the states of this storage can be described by
elements of
$\Cal B(\Delta)$.
 But if
$\Delta$
is a set of qubits
forming a quantum storage of a quantum computer, then mixed states
of this storage can be described by elements of
${\bold D}(\Delta)$.
 Of course, pure states are characterized by elements of
$\Cal H(\Delta)$
(up to a nonzero number coefficient).

 A linearly ordered subset
$X$
of
$\Delta$
is called {\it register}.
 Pure states of
$X$
correspond to elements of
$\Cal H(X)$;
general mixed states of
$X$
correspond to elements of
${\bold D}(X)$,
i.e., density operators in
$\Cal H(X)$.
 In
particular, each qubit is a register.
 In this case
$X$
consists of a single element and
$\Cal H(X)$
is two-dimensional.
 If
$X$
consists of
$n$
qubits, then
$\Cal H(X)$
is a
tensor product of
$n$
two-dimensional Hilbert spaces corresponding
to each qubit.

 Each qubit has two basic states denoted by
$\vert 0\rangle$
and
$\vert 1\rangle$
(Dirac's notation is used).
 States of
a quantum storage (or its register) are called {\it classical
states} if they are tensor products of these basic states
corresponding to each qubit.
 It is assumed that the quantum
storage can be prepared (initiated) in an arbitrary classical
state.

 Any unitary operator
$U\in {\bold U}(X)$
defines a transformation
$S\mapsto S^U$
on the set
${\bold D}(X)$
of all states of the
register
$X$
by the formula
$S^U=USU^{-1}$.
 We shall say that
a unitary operator
$U\in {\bold U}(\Delta)$
is {\it concentrated
on a register}
$X\subset \Delta$,
if
$U$
can be represented in the form
$U=U_X\otimes id_Z$,
where
$U_X\in{\bold U}(X)$,
$Z=\Delta\backslash X$,
and
$id_Z$
is the identity operator in
$\Cal H(Z)$.
 Respectively, we
shall also consider any unitary operator
$U_X\in {\bold U}(X)$
as an
operator of the form
$U_X\otimes id_Z$
belonging to
${\bold U}(\Delta)$,
where
$Z=\Delta\backslash X$.
 We shall say that the register
$X$
is a {\it support} of
$U\in{\bold U}(\Delta)$
and denote it by
$\supp (U)$,
if
$X$
is
the minimal register on which~$U$
is concentrated.

 A quantum computer performs unitary transformations in
$\Cal H(\Delta)$.
 It is assumed that in one step an elementary unitary transformation
can be made and there is a fixed collection (basis) of such
unitary operators which are called {\it logic gates}, or simply
{\it gates}.
 It is also assumed that every gate has a short
support (usually consisting of one or two qubits).
 Combinations
of these gates define {\it quantum circuits}.

 Every bijection
$\sigma$
of the set of all classical states
${\Cal B}(X)$
onto itself leads to a permutation of the elements of
the corresponding orthonormal basis in
${\Cal H}(X)$
and generates
a unitary operator
$\widehat\sigma\in {\bold U}(X)$.
 Operators of this type are called {\it classical operators}
({\it transformations}).

\example{Example 1} For a register
$\{ x\}$
consisting of a single qubit
$x$
the permutation
$\vert 0\rangle\mapsto\vert 1\rangle$,
$\vert 1\rangle\mapsto\vert 0\rangle$
defines the so-called {\it
negation} operator (or NOT operator) denoted by
$\lnot_x$.
\endexample

\example{Example 2} Another important example is the so-called
controlled NOT (or CNOT) operator, see, e.g., \cite{7}.
 For a two-bit
register
$X=\{x,y\}$
this operator is induced by the bijection
$\tau :\vert x,y\rangle\mapsto\vert x,x\oplus y\rangle$,
where
``$\oplus$'' denotes addition modulo 2.
 For classical states this
bijection
$\tau$
allows to copy the content of one bit into another,
provided the second bit is empty.
 Of course, a similar operator can
be defined for a pair of arbitrary registers of the same size
by applying
$\tau$
to each pair of bits.
\endexample

 According to the above, the operators described in these
examples can be treated as unitary operators belonging to
${\bold U}(\Delta)$.

 Thus, for the standard model, (mixed) states of a finite storage
$\Delta$
are defined by density operators belonging to
${\bold D}(\Delta)$,
whereas elementary operations (gates) are
defined by a fixed collection of unitary operators concentrated
on short registers.
 In the framework of this model, the execution of an algorithm
starts from a preparation of the storage
$\Delta$
in a
classical state.
 Then a sequence of unitary quantum gates is applied.
 Finally, a measurement operation (which is a specific type of
interaction between the quantum computer and an external
physical device) is performed.
 The result of this measurement
operation is a classical state of the register.
 For the corresponding
details, see, e.g.,~\cite{11--13}.

\ahead{2.
 Quantum computer media model}\endahead

 There are rather many different paradigms and models for quantum
computer systems, see e.g. \cite{2, 11--19}.
 We shall say that
a computer medium including a system of parallel quantum
computers (processors) and classical components is a {\it quantum
computer medium} (briefly QCM).
 We shall consider a
version of this model convenient for our aims.
 This QCM has a
storage
$\Delta$
which is a set of large ensembles called
$q$-{\it ensembles}.
 Roughly speaking, any
$q$-ensemble can be
regarded as a flow of independent qubits, whereas operations with
$q$-ensembles can be treated as actions independently affecting
each qubit in the same way under the same conditions (of course,
the number of qubits in a
$q$-ensemble is finite
but large enough).
 There is a similar situation, say, in
so-called bulk quantum computation, where one can manipulate a
large number of indistinguishable quantum computers by parallel
unitary
operations; see, e.g., \cite{15} for details and implementations
using nuclear magnetic resonance.

 Denote by
$\Delta$
the set of all
$q$-ensembles forming the storage
of our QCM and denote by
$\widetilde \Delta$
the set of all
qubits belonging to this storage.
 Any
$q$-ensemble
$x\in\Delta$
forms a subset
$\widetilde\Delta_x$
in
$\widetilde\Delta$.
 We say that its {\it state}
$\widetilde S_x\in{\bold D} (\widetilde\Delta_x)$
{\it is admissible}, if
$\widetilde S_x$
is a tensor power of a state
$S\in{\bold D}(\{a\})$,
where
$a\in\widetilde\Delta_x$,
so that there is a one-to-one
correspondence between the set of all admissible
$q$-ensemble states and the set of all states for each qubit belonging
to this
$q$-ensemble.
 Every mixed state
$\widetilde{S}\in {\bold D}(\widetilde\Delta)$
can be restricted to any
$q$-ensemble (by the partial trace formula, see, e.g., \cite{11, 12}
and
below).
 If all such restrictions are admissible, then we say that
the {\it state}
$\widetilde{S}$
{\it is admissible}.
 Denote by
${\bold D}^*(\Delta)$
the set of all admissible states of our QCM.
 We shall say that a {\it unitary operator}
$U\in{\bold U}(\widetilde\Delta)$
{\it is admissible}, if
${\bold D}^*(\Delta)$
is invariant under the action of this operator.

 It is clear that the corresponding standard quantum computer model
can be embedded into the QCM model, so that
${\bold D}(\Delta)$
and
${\bold U}(\Delta)$
correspond to the sets of admissible states and
admissible unitary operators respectively.
 It is assumed that the set
of all classical states in the standard model can be identified with
the corresponding set of states in the QCM model.
 So every algorithm
implemented in the framework of the standard model can be transferred
to the QCM model.

 However, for the QCM case, it is possible to construct a cloning
(copying) operation which transfers any
$q$-ensemble in an
admissible state to a pair
of
$q$-ensembles such that each of their qubits has the same
state.
 This copying operation may be treated as the division of the
initial
$q$-ensemble into large parts.
 Note that in the framework of the standard model, perfect cloning is
impossible: an unknown quantum state can not be cloned
(unless this state is already known, i.e., there exists a
classical information which specifies it).
 However, it is
possible to make approximate copies.
 For details see, e.g.,
\cite{13, 20--23}.

\ahead{3.
 Modeling the arithmetic of real numbers}\endahead

 Thus, in the framework of the QCM model, it is possible to
implement the standard model with mixed states and a cloning
operation.
 Below we need to obtain a collection of qubits prepared in
identical classical states and to manipulate these copies by
parallel unitary operations.
 So, for the sake of simplicity, we shall consider the
standard model extended by this cloning operation.
 This operation
is not a quantum unitary operation.
 However, we shall include
it in quantum circuits (a similar trick was used in
\cite{13} with respect to measurement operations).
 For this quantum
computer an implementation of the real number arithmetic is
presented below.
 Moreover, for the arithmetical operations, the
execution time does not depend on the complexity (in
the usual sense) of operands.
 In particular, for the function
$n\mapsto a^n$,
where
$n\in {\bold N}$
and
$a$
is an arbitrary
real number, it is possible to get a polynomial algorithm of its
calculation (with respect to the size of the number
$n$,
i.e.,
$\log n$)
using the well-known standard trick:
$a\mapsto a^2\mapsto a^4=(a^2)\cdot(a^2)$
etc.

 Using the notation introduced above in Sec.~1, denote
by
${\Cal B}(n)$
the set
${\Cal B}(\{ 1,\cdots,n\})$
and by
${\Cal H}(n)$
the Hilbert space
${\Cal H} (\{1,\cdots,n\})$.
 Denote by
$\vert 0\rangle$
and
$\vert 1\rangle$
elements of
${\Cal B}(1)$
and by
$\vert\alpha_1,\cdots,\alpha_n\rangle$
elements
of
${\Cal B}(n)$,
where
$\vert\alpha_i\rangle\in{\Cal B}(1)$.
 Using Dirac's bra/ket notation, denote by
$\vert x\rangle$
elements
of the Hilbert space
${\Cal H}(X)$
(ket-vectors), and by
$\langle x\vert y\rangle$
the scalar product of the vectors
$\vert x\rangle,\vert y\rangle\in {\Cal H}(X)$.
 Any
bra-vector
$\langle x\vert$
corresponds to the linear functional
$y\mapsto\langle x\vert y\rangle$
on
${\Cal H}(X)$,
whereas the notation
$\vert a\rangle\langle b\vert$
corresponds to the linear
operator
$\vert x\rangle\mapsto\langle b,x\rangle\vert a\rangle$.

 We have assumed that each register is linearly
ordered; therefore
${\Cal B}(X)$
and
${\Cal H}(X)$
can be naturally identified with
${\Cal B}(n)$
and
${\Cal H}(n)$,
where
$n$
is the length of the
register
$X$
(i.e., number of elements of
$X$).

 Let
$S$
be a (mixed) state of the storage
$\Delta$
and let
$X\subset\Delta$
be an arbitrary register.
 The {\it restriction of the state}
$S$
{\it to the register}
$X$
is defined by the partial trace formula
$$ %nmbr
S\mapsto S(X)=\Tr_Z(S)\in {\bold D}(X),
\qq\text{where}\q
Z=\Delta\backslash X,
$$ %endnmbr
(see, e.g., \cite{11, 12}) and we shall say that
$S(X)$
is the state of~$X$.

 Any state of a one-point register
$\{x\}$,
i.e., of a qubit, is defined
by the corresponding density matrix
$S=(S_{ij})$
with respect to
the basis
${\Cal B}(\{ x\})=\{\vert 0\rangle,\vert 1\rangle\}$.
 Here the matrix element
$S_{00}$
is equal to the probability
that the measured value of the qubit is
$\vert 0\rangle$.
 Similarly, the probability that the measured value of the qubit is
$\vert 1\rangle$
coincides with
$S_{11}$.
 Any classical state of
$X$,
i.e., any element
$f\in{\Cal B}(X)$,
corresponds to the operator
$\otm_{x\in X}\vert f(x)\rangle\langle f(x)\vert$.
 In particular, we set
$$
\0_X = \otm_{x\in X}\vert 0\rangle\langle 0\vert,
\qquad
\1_X = \otm_{x\in X}\vert 1\rangle\langle 1\vert.
$$
 These operators correspond to the Boolean functions on
$X$
(i.e., classical states) which are identically equal to 0 or 1.

 Consider a partition
$X=\cuplim_{i=1}^n X_i$
of an arbitrary register
$X$
into disjoint registers
$X_i$.
 It is clear that in this case the space
${\Cal H}(X)$
can be decomposed in the form of the tensor product
$$ %nmbr
{\Cal H}(X)=\otm_{i=1}^n{\Cal H}(X_i)
$$ %endnmbr
of the spaces
${\Cal H}(X_i)$.
 Therefore, for any collection of operators
$A_i\in {\bold L}(X_i)$,
their tensor product
$$ %nmbr
A=\otm_{i=1}^n A_i\in{\bold L}(X)
$$ %endnmbr
exists.
 We shall say that this operator
$A$
is {\it decomposable with respect
to the partition}
$X=\cuplim_{i=1}^n X_i$.
 Note that the tensor product
of density operators is a density operator,
and in the same way the tensor
product of unitary operators is a unitary operator.
 We say that a state
(i.e., an element of
${\bold D}(X)\subset{\bold L}(X)$)
is (simply) {\it decomposable}, if it is decomposable with respect to
the
partition of
$X$
into its points, i.e., into one-qubit registers.
 Note that
every classical state is decomposable.

 Let
$A=A_1\cup A_2$
be a partition of a register
$A$
into disjoint
registers
$A_1$
and
$A_2$,
$X_1\subset A_1$
and
$X_2\subset A_2$.
 It is easy to check that if a state
$S\in {\bold D}(A)$
is decomposable with respect to the partition
$A=A_1\cup A_2$,
then the restriction of this state to the register
$X=X_1\cup X_2$
is decomposable with respect to the partition
$X=X_1\cup X_2$.
 We say that a state
$S\in {\bold D}(\Delta)$
is {\it decomposable
with respect to a register}
$X\subset\Delta$,
if its restriction
$S(X)$
to
$X$
is decomposable.

 We say that a register
$F$
is {\it free with respect to a state}
$S$
and that this state
$S$
{\it is free with respect to the register}
$F$
if
$$ %nmbr
S=S'\otimes \0_F,
\qq\text{where}\q
S'\in {\bold D}(\Delta\backslash F).
$$ %endnmbr
 If
$U\in {\bold U}(\Delta)$
and
$F$
is free with respect to
$S$,
then
the register
$F\backslash \supp (U)$
is free with respect to the state
$S^U=USU^{-1}$.

 Now we can discuss our implementation of the arithmetic of real
numbers.
 Let
$S$
be a state of a register
$X$;
then denote by
$S(x)$
the restriction of
$S$
to a qubit
$x\in X$.
 Suppose
$$ %nmbr
X=\{ x_1,x_2,x_3,x_4\}
$$ %endnmbr
is a register
consisting of four qubits,
$S$
is a decomposable state of
$X$,
$S(x)_{ij}$
is the corresponding density matrix.

 We shall say that any decomposable state
$S$
of
$X=\{x_1,x_2,x_3,x_4\}$
represents the following real number:
$$
r(S)=
{{(S(x_1))_{11}-(S(x_2))_{11}}\over{(S(x_3))_{11}-(S(x_4))_{11}}}.
\tag{1}
$$
 Of course, different states of
$X$
may represent the same real number.
 In particular, every real number can be represented by a pure state.
 We
say that real numbers represented in the form (1) are {\it numbers
of\/}
{\bf real}4 {\it type}.

 Any arithmetical operation
$\circledast$
(e.g. multiplication or addition)
is implemented by a circuit
${\Cal U}$.
 Suppose that
$F$
is a free
storage and numbers are located in the disjoint registers
$A$
and
$B$.
 Assume that
$S$
is an initial state which is free with respect to
$F$
and decomposable with respect to
$A\cup B$.
 The corresponding circuit
$\Cal U$
transfers
$S$
to a state
$\widetilde S$
such that the
restriction
$\widetilde S(A)$
of
$S$
to the register
$A$
is decomposable
and
$$ %nmbr
r(\widetilde S(A))=r(S(A))\circledast r(S(B)).
$$ %endnmbr
 The circuit
$\Cal U$
is a fixed finite combination of unitary operators belonging
to a fixed collection of gates.
 It is natural to say that this collection is a set of instructions
for the corresponding arithmetical processor.

 Along with the numbers of {\bf real}4 type, we shall also
consider numbers of {\bf real}1 type and {\bf real}2 type.
 Any
state
$S(x)$
of a one-qubit register
$\{ x\}$
represents the following
real {\it number of\/} {\bf real}1 type:
$$
r(S(x))=(S(x))_{11},\tag{2}
$$
so
$0\leq r(S(x))\leq 1$.

 Let
$B=\{b^+,b^-\}$
be a two-qubit register,
$S=S(B)$
its decomposable
state; then we say that
$S(B)$
represents the following real number
of {\bf real}2 type:
$$
r(S(B))=(S(b^+))_{11}-(S(b^-))_{11}=r(S(b^+))-r(S(b^-)).\tag{3}
$$
 Of course,
$-1\leq r(S(B))\leq 1$
and every such number can be represented
by a pure state.
 It is clear that every number
$r$
of {\bf real}4
type can be treated as a pair of numbers
$(r',r'')$
of {\bf real}2 type,
where
$r=r'/r''$.

 Let
$S$
be a state of a one-qubit register; then we say that
$S$
is
{\it diagonal} if the corresponding matrix
$S_{ij}$
($i,j=1,2$)
is
diagonal.
 Suppose
$S$
is a state of an arbitrary register; then we say
that the {\it state}
$S$
is {\it diagonal} if it is decomposable and
its restriction to every element (qubit) of the register is diagonal.
 We shall say that {\it states of one-qubit registers are equivalent},
if their density matrices have the same diagonal elements, i.e.,
represent the same number of {\bf real}1 type.
 We say that {\it
states of registers of the same length are equivalent} if the
restrictions of these states to the corresponding components (qubits)
are equivalent.

 Let us describe an operation which transfers states of qubits to
equivalent diagonal states using a free storage.
 If
$S$
is a state
of a two-qubit register
$X=\{ a,b\}$
and
$\{b\}$
belongs to the free
storage, then this operation transfers
$S$
to a diagonal state
$S'$
such that
$S'(a)$,
$S'(b)$,
and
$S(a)$
are equivalent.
 To this
end the CNOT operation (described in the example 2 above) can be used.
 It is elementary to check that the following proposition is true.

\proclaim{Proposition 1} Let
$S=S(a)\otimes\0_{\{ b\}}$
be a state of
a register
$X=\{a,b\}$,
$S^{\tau}=USU^{-1}$,
where
$U=\hat\tau$
is the
classical CNOT operator described in Example~\rom2.
 Then
$$ %nmbr
S^{\tau}(a)=S^{\tau}(b)=(\delta_{ij} (S(a))_{ij}),
$$ %endnmbr
where
$i,j =1,2$
and
$\delta_{ij}$
is the Kronecker symbol. \endproclaim

\ahead{4.
 Implementation of arithmetical operations}\endahead

\chead
4.1
\endchead
 We shall describe a set of instructions (i.e., unitary
operators) ensuring the implementation of arithmetical operations.
 Finally, we shall describe
two elementary operations for all numbers
$x$,
$y$
of {\bf real}1 type.
 Set
$$
\align
\sigma_1 & :\; x, y \mapsto {{x+y}\over 2},\tag{4}\\
\sigma_2 & :\; x, y \mapsto 1-(x+y)+ 2xy.\tag{5}
\endalign
$$

 The corresponding unitary operators act on two-qubit registers
$X=\{ a,b\}$,
i.e., on the space
${\Cal H}(X)={\Cal H}(2)$.

 A direct verification shows that if an input state
$S$
is decomposable with respect to
$X=\{ a,b\}$,
i.e.,
$S(X)=S(a)\otimes S(b)$,
then the operation
$\sigma_2$
can be implemented by a classical operator;
this operator is generated by the following permutation of elements
of the standard orthonormal basis in
${\Cal H}(2)$:
$$
\vert 1,1\rangle\mapsto\vert 1,1\rangle,\qquad
\vert 1,0\rangle\mapsto\vert 0,0\rangle,\qquad
\vert 0,0\rangle\mapsto\vert 0,1\rangle,\qquad
\vert 0,1\rangle\mapsto\vert 1,0\rangle.
$$

 Similarly, let
$S$
be an input state decomposable with respect to
$X=\{ a,b\}$
and diagonal for each its qubit.
 In this case the
operation
$\sigma_1$
is implemented by the unitary operator
$U$
which acts on the standard basis in
${\Cal H}(2)$
by the following way:
$$
\align
\vert 1,1\rangle & \mapsto\vert 1,1\rangle,\qquad
\vert 0,0\rangle\mapsto\vert 0,0\rangle,\\
\vert 1,0\rangle & \mapsto\lambda(\vert 1,0\rangle+\vert
0,1\rangle),\qquad
\vert 0,1\rangle \mapsto \lambda(-\vert 1,0\rangle+\vert 0,1\rangle),
\endalign
$$
where
$\lambda=1/\sqrt{2}$.
 The operation
$\sigma_1$
is implemented by the
operator
$U$
only in the case of diagonal state
$S$.
 However,
the CNOT operator
$\widehat\tau$
(see Example 2 and Proposition 1 above)
with a qubit of a free storage transfers the qubit in the state
$S$
to
a diagonal state
$S'$
such that
$r(S')=r(S)$.
 Therefore, combining
$U$
with
$\widehat\tau$,
we obtain an implementation
of the operation
$\sigma_1$.

 Combining the operations
$\sigma_1$
and
$\sigma_2$
with cloning
operations, it is possible to compute the following function
$$
\align
\mu_1(x,y) & = \sigma_1(\sigma_1(\sigma_2(x,y),0),\sigma(x,y))\\ & =
((1-(x+y)+2xy +0)/2+(x+y)/2)/2= xy/2+1/4.
\endalign
$$

 Thus, we have proved the following
\proclaim{Proposition 2} For all numbers of the {\bf real}{\rm 1}
type the operations of arithmetic mean
$x,y \mapsto (x+y)/2$
and displaced multiplication
$x, y \mapsto xy/2 +1/4$
can be
implemented by fixed quantum circuits\footnote{This means that every
circuit is a fixed combination of gates (instructions).}.\endproclaim
\smallskip

\chead
4.2
\endchead
 Let us show now that for all numbers of
{\bf real}2 type a similar proposition is valid.

\proclaim{Proposition 3} For all numbers of {\bf real}{\rm 2}
type the operations of arithmetic mean
$x, y \mapsto (x+y)/2$
and quasimultiplication
$x, y \mapsto xy/4$
can be
implemented by fixed quantum circuits.\endproclaim

 Recall that any number
$z$
of {\bf real}2 type can be
represented as the difference
$z^+-z^-$
of numbers of
{\bf real}1 type.

 By abuse of language, we denote by
$\sigma$
the operation of arithmetic
mean and by
$\mu_2$
the operation of quasimultiplication from the
proposition 3.

 An implementation of the operations stated in Proposition 3
can be given by the following formulas:
$$
\align
\sigma^+(x,y) &= \sigma(x^+,y^+)=(x^++y^+)/2,\\
\sigma^-(x,y) &= \sigma(x^-,y^-)=(x^-+y^-)/2;\\
\mu^+(x,y) &= \sigma(\mu_1(x^+,y^+),\mu_1(x^-,y^-))=
\sigma(x^+y^+/2+1/4,x^-y^-/2+1/4)\\ & =(x^+y^++x^-y^-)/4+1/4,\\
\mu^-(x,y) &= \sigma(\mu_1(x^+,y^-),\mu_1(x^-,y^+))=
\sigma(x^+y^-/2+1/4,x^-y^+/2+1/4)\\ & =(x^+y^-+x^-y^+)/4+1/4.
\endalign
$$
 Indeed,
$$ %nmbr
\gather
\sigma(x,y) = \sigma^+(x,y)-\sigma^-(x,y)=(x+y)/2,
\\
\mu_2(x,y) = \mu^+(x,y)-\mu^-(x,y)=(x^+y^++x^-y^--x^+y^--x^-y^+)/4=
xy/4,
\endgather
$$ %endnmbr
as was to be proved.
 Of course, this calculation needs copying
operations.
\bigskip

\chead
4.3
\endchead
 From Proposition 2 and 3 we can easily deduce the
following

\proclaim{Theorem} For all numbers of {\bf real}{\rm 4} type,
the operations of addition, multiplication, subtraction, and division
can
be implemented by fixed quantum circuits.
\endproclaim

 Recall that any number
$z$
of {\bf real}4 type can be represented
in the form
$z'/z''$,
where
$z'$
and
$z''$
are numbers of
{\bf real}2 type.
 The operation of multiplication is given by
the following formulas:
$$
(xy)'=\mu_2(x',y'),\qquad (xy)''=\mu_2(x'',y'').
$$
 The arithmetic mean
$(x+y)/2$
is given by the formulas
$$
((x+y)/2)'=\sigma(\mu_2(x',y''),\mu_2(x'',y')),\qquad
((x+y)/2)''=\mu_2(x'',y'').
$$
 Finally, the sum
$x+y$
can be obtained by the multiplication of
the numbers
$(x+y)/2$
and 2.
 Note that the number 2 can be easily
implemented as a number of {\bf real}4 type.

 From the basic formula (1) it is clear that for any real number
$r$
and its representation in the form (1) we can easily construct
the corresponding representations for the numbers
$-r$
and
$r^{-1}$;
so the operations of subtraction and division can also
be implemented
by fixed quantum circuits.

\remark{Remark} Note that our exact definitions and constructions
of the operations are not stable with respect to small
perturbations.
 However, all the elementary operations are
continuous and in practice we shall deal with approximate
values and operations, errors, etc. (as usual for
calculations with real numbers).
 So we need to examine the
corresponding methods for fault-tolerant calculations.
 This will be the subject of our subsequent publications.
\endremark

\ahead
 ACKNOWLEDGMENTS
\endahead

The authors are grateful to S.~L.~Braunstein and P.~Benoiff for useful
comments and references.

This research was supported
by the Russian Foundation for Basic Research
under grants no.~96-01-01544 and 99-01-01198
and
by the Erwin Schr\"odinger
 International Institute for Mathematical Physics (Vienna).

%%%%%%%%% R E F E R E N C E S (to be checked
\par\removelastskip\penalty-200\medskip

\Refs

\item{1.}
S.~L.~Braunstein,
``Error correction
for continuous quantum variables,''
{\it Phys. Rev. Lett.},
{\bf 80}
(1998),
no.~18,
4084--4087.

\item{2.}
S.~Lloyd and S.~L.~Braunstein,
``Quantum computation
over continuous variables,''
{\it Phys. Rev. Lett.},
{\bf 82}
(1999),
no.~8,
1784--1787.

\item{3.}
P.~Benioff,
``The computer as a physical system: a microscopic quantum mechanical
Hamiltonian model of computers as represented
by Turing Machines,''
{\it J. Statist. Phys.},
{\bf 22}
(1980),
no.~5,
563--591.

\item{4.}
P.~Benioff,
``Quantum mechanical Hamiltonian models of Turing machines,''
{\it J. Statist. Phys.},
{\bf 29}
(1982),
515--546.

\item{5.}
%Ю.~И.~Mанин,
Yu.~I.~Manin,
{\it The Computable and Incomputable\/}
[in Russian],
Soviet Radio,
Moscow,
1980.

\item{6.}
R.~P.~Feynman,
``Simulating physics
with computers,''
{\it Int. J. Theor. Phys.},
{\bf 21}
(1982),
no.~6/7,
467--488.

\item{7.}
R.~P.~Feynman,
``Quantum Mechanical Computers,''
{\it Optic News},
{\bf 11}
(1985),
11--20,
{\it Foundations of Physics},
{\bf 16}
(1986),
no.~6,
507--531.

\item{8.}
A.~Peres,
``Reversible logic and quantum computers,''
{\it Phys. Rev.~A},
{\bf 32}
(1985),
3266--3276.

\item{9.}
D.~Deutsch,
``Quantum theory, the Church--Turing principle and the universal
quantum
computer,''
{\it Proc. Roy. Soc. London. Ser.~A},
{\bf 400}
(1985),
97--117.

\item{10.}
D.~Deutsch,
``Quantum computational networks,''
{\it Proc. Roy. Soc. London. Ser.~A},
{\bf 425}
(1989),
73--90.

\item{11.}
A.~Yu.~Kitaev,
``Quantum measurements and the abelian stabilizer problem,''
in: {\it E-print {\tt quant-ph/9511026}},
1995.

\item{12.}
%~A.~Ю.~Китаев,
A.~Yu.~Kitaev,
``Quantum computations: algorithms and error correction,''
{\it Uspekhi Mat. Nauk\/} [{\it Russian Math. Surveys\/}],
{\bf 52}
(1997),
no.~6,
53--112.

\item{13.}
D.~Aharonov, A.~Kitaev, and N.~Nisan,
``Quantum circuits
with mixed states,''
in: {\it E-print {\tt quant-ph/9806029}},
1998.

\item{14.}
A.~Steane,
``Quantum computing,''
{\it Rep.~Prog. Phys.},
{\bf 61}
(1998),
117--173.

\item{15.}
E.~Knill, I.~Chuang, and R.~Laflame,
``Effective pure states
for bulk quantum computation,''
in: {\it E-print {\tt quant-ph/9706053}},
1997.

\item{16.}
S.~C.~Benjamin and N.~F.~Johnson,
``Structures
for data processing
in the quantum regime,''
in: {\it E-print {\tt quant-ph/9802127}},
1998.

\item{17.}
A.~K.~Ekert,
``Distributed quantum computations
over noisy channels,''
in: {\it E-print} {\tt quant-ph/\allowlinebreak9803017},
1998.

\item{18.}
Yu.~Ozhigov,
``Quantum computers speed up classical
with probability zero,''
in: {\it E-print {\tt quant-ph/9803064}},
1998.

\item{19.}
C.~Moore and M.~Nilsson,
``Parallel quantum computation and quantum codes,''
in: {\it E-print {\tt quant-ph/9808027}},
1998.

\item{20.}
H.~Barnum,
``Noncommuting mixed states cannot be broadcast,''
{\it Phys. Rev. Lett.},
{\bf 76}
(1996),
2818--2821.

\item{21.}
V.~Buzek and M.~Hillary,
``Quantum copying beyond the no-cloning theorem,''
{\it Phys. Rev. Lett.~A},
{\bf 54}
(1996),
1844--1862.

\item{22.}
N.~Gisin and S.~Massar,
``Optimal quantum cloning machines,''
{\it Phys. Rev. Lett.},
{\bf 79}
(1997),
2153--2156.

\item{23.}
D.~Bruss, A.~Ekert, and C.~Macchiavello,
``Optimal universal quantum cloning and state estimation,''
{\it Phys. Rev. Lett.},
{\bf 81}
(1998),
2598--2601.

\endRefs
%%%%%%%%%

\authoraddress%
{(G.~L.~Litvinov, G.~B.~Shpiz) International Sophus Lie Center}
\authoremail%
{islc\@ dol.ru}

\authoraddress%
{(V.~P.~Maslov) M.~V.~Lomonosov Moscow State University}

%\transl{G.~L.~Litvinov} %translator's name
%%%%%%%%%

\enddocument